\documentclass{kapproc} 

\usepackage{graphicx}

\upperandlowercase

\kluwerbib 

\begin{document}

\articletitle{The starburst phenomenon from the optical/near-IR perspective}


\author{Nils Bergvall,\altaffilmark{1} Thomas Marquart,\altaffilmark{1} G\"oran 
\"Ostlin,\altaffilmark{2} Erik Zackrisson\altaffilmark{1}}

\altaffiltext{1}{Uppsala Astronomical Observatory\\
Box 515, SE-751 20 Uppsala, Sweden}
\email{nils.bergvall@astro.uu.se, thomas.marquart@astro.uu.se, ez@astro.uu.se}


\altaffiltext{2}{Stockholm Observatory\\
AlbaNova University Center, Roslagstullsbacken 21,
SE-106 91 Stockholm, Sweden}
\email{ostlin@astro.su.se}


\begin{abstract}
The optical/near-IR stellar continuum carries unique information about the 
stellar population in a galaxy, its mass function and star-formation history. 
Star-forming regions display rich emission-line spectra from which we can derive 
the dust and gas distribution, map velocity fields, metallicities and young 
massive stars and locate shocks and stellar winds. All this information is very 
useful in the dissection of the starburst phenomenon. We discuss a few of the 
advantages and limitations of observations in the optical/near-IR region and 
focus on some results. Special attention is given to the role of interactions 
and mergers and observations of the relatively dust-free starburst dwarfs. In 
the future we expect new and refined diagnostic tools to provide us with more 
detailed information about the IMF, strength and duration of the burst and its 
triggering mechanisms.  
\end{abstract}


\section{Introduction}
\begin{figure}[h]
\includegraphics[width=0.65\textwidth]{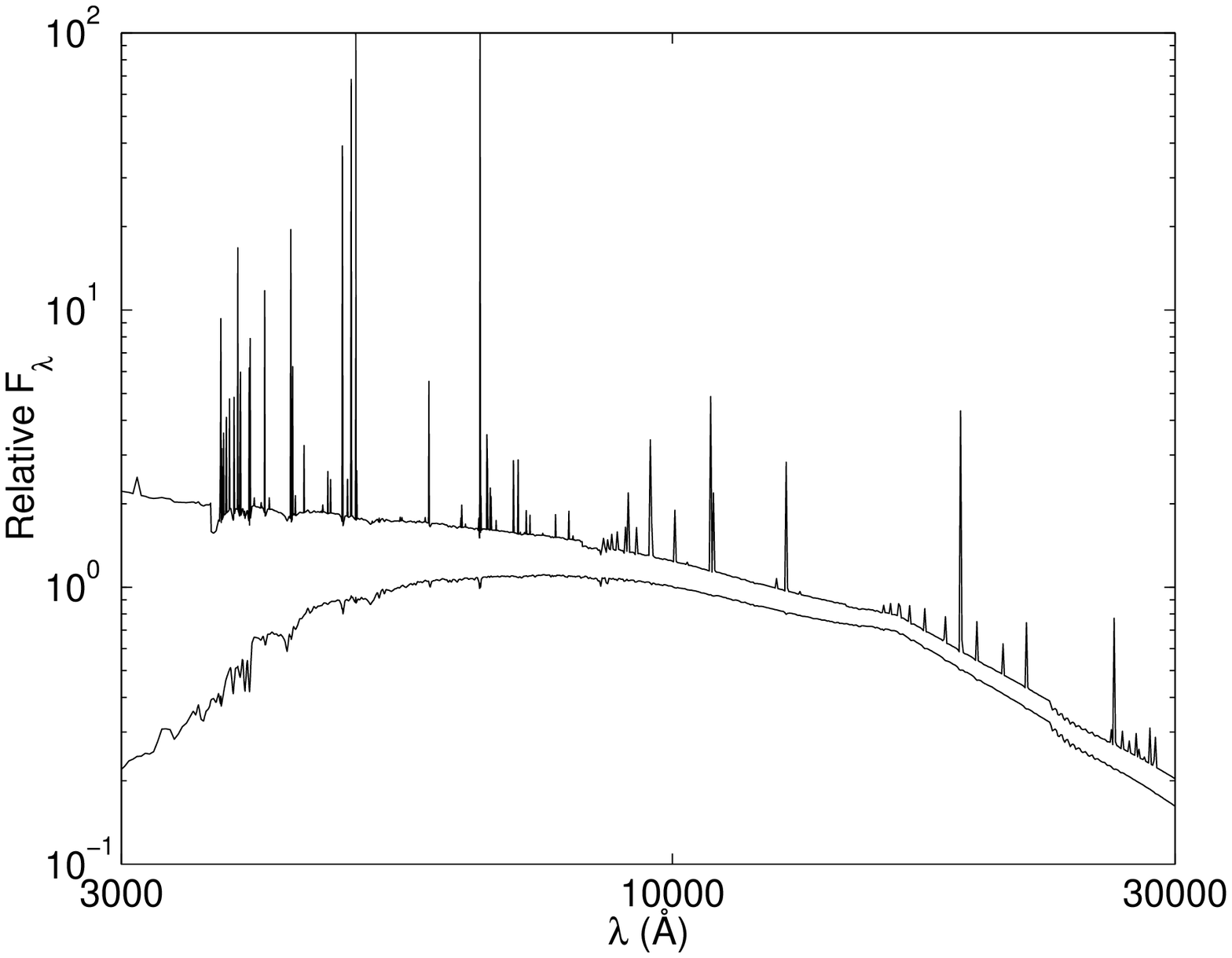}
\narrowcaption[]{Synthetic spectrum (Zackrisson et al.~2001) of a starburst in 
an old galaxy. The first burst (lower spectrum) started 10 Gyr ago and had a 
duration of 100 Myr. The young starburst has a constant SFR and is 10 Myr old. 
The mass (including stellar remnants) of the old population is twice that of the 
young population. Both populations have 5\% solar metallicity and a Salpeter 
IMF.}
\label{specfig}
\end{figure}

Optical/near-IR broadband photometry of a starburst galaxy gives a first 
indication of burst strength, age and distribution of the young and old 
populations and their basic morphological structure parameters. Model based 
spectrophotometric tools are provided for more detailed analysis. A rich set of 
emission lines are used for analysis of kinematics, chemical abundance, shocks, 
stellar upper mass limit and distribution of dust and molecular gas. Absorption 
line indices provide estimates of age and IMF of the evolved population. 
Fig.~\ref{specfig} shows a synthetic spectrum a mixture of a young and old 
population with a mass ratio 2:1. A general review of the diagnostic tools and 
the limitations of the photoionization models used in the analysis is discussed 
by Schaerer (2001). 

 Heavy dust obscuration, in particular in LIRGs and ULIRGs, has been a problem 
in the optical/near-IR. Here we will therefore focus on starbursts in 
low-extinction regions, notably starburst dwarf galaxies. First, however, we 
will discuss a widely debated issue where optical data originally had a strong 
impact, namely the importance of gravitational interactions as a starburst 
triggering mechanism.

\section{Starbursts and tidal interaction}

It is clear from the properties of ULIRGs that mergers are required to trigger 
major starbursts. But is it a sufficient requirement? How often do mergers and 
close encounters generate starbursts? To answer this question it is common to 
compare two galaxy samples - interacting/merging galaxies (IGs) or pairs, and 
non-interacting galaxies (NIGs). A problem with the comparison is that NIGs and 
IGs have evolved in different environments where e.g. mergers, ram pressure, 
harassment and gas infall have different influence.
Integrated broadband photometry and H$\alpha$ emission are the most widely used 
tools in this context. In the classical paper by Larson \& Tinsley (1978) the 
authors claim, based on UBV data, that interactions frequently trigger a major 
SF increase involving as much as 5\% of the total mass. Many follow-up studies 
seem to confirm the result but are often influenced by strong selection effects, 
non-matching morphological type distribution NIGs/IGs and are focusing on the 
most dramatic cases. Studies based on more well constrained samples (Bergvall et 
al.~2002, Brosch et al.~2004) do not confirm these results but find that tidal 
interactions have an insignificant influence on the SF history of galaxies in 
the local universe. There seems to be an agreement however, of a correlation 
between interaction and increased SF within the central kpc (first discussed by 
Keel et al.~1985). {\it Galaxy pairs} with small separations show similar trends 
as seen in H$\alpha$ (Barton et al.~2000, Lambas et al.~2003, Nikolic et 
al.~2004). The mean increase is in both cases is quite moderate however, and few 
cases are qualified to be called 'nuclear starbursts'. Bergvall et al.~(2000) 
and Varela et al.~(2004) find that masses of perturbed galaxies are higher than 
NIGs of similar morphology indicating that they experience mergers more 
frequently. This may lead to a steady inflow of gas that can explain part of the 
increased SF in the centre. Varela et al.~also find a {\it higher frequency of 
bars in disturbed systems}, in accordance with related studies in the past (see 
Knapen 2004). Bars are known to generate mass inflows. Thus it is not clear what 
is the main triggering mechanism of the central increase in SF. The conclusion 
must be that there is {\it no strong support that tidal interactions generate 
starburst activity that significantly affects the SF history of galaxies in the 
local universe}. Estimates give room for major starbursts among less than a few 
\% of the IGs.

\section{Blue compact galaxies}

\begin{figure}[h]
\includegraphics[width=0.6\textwidth]{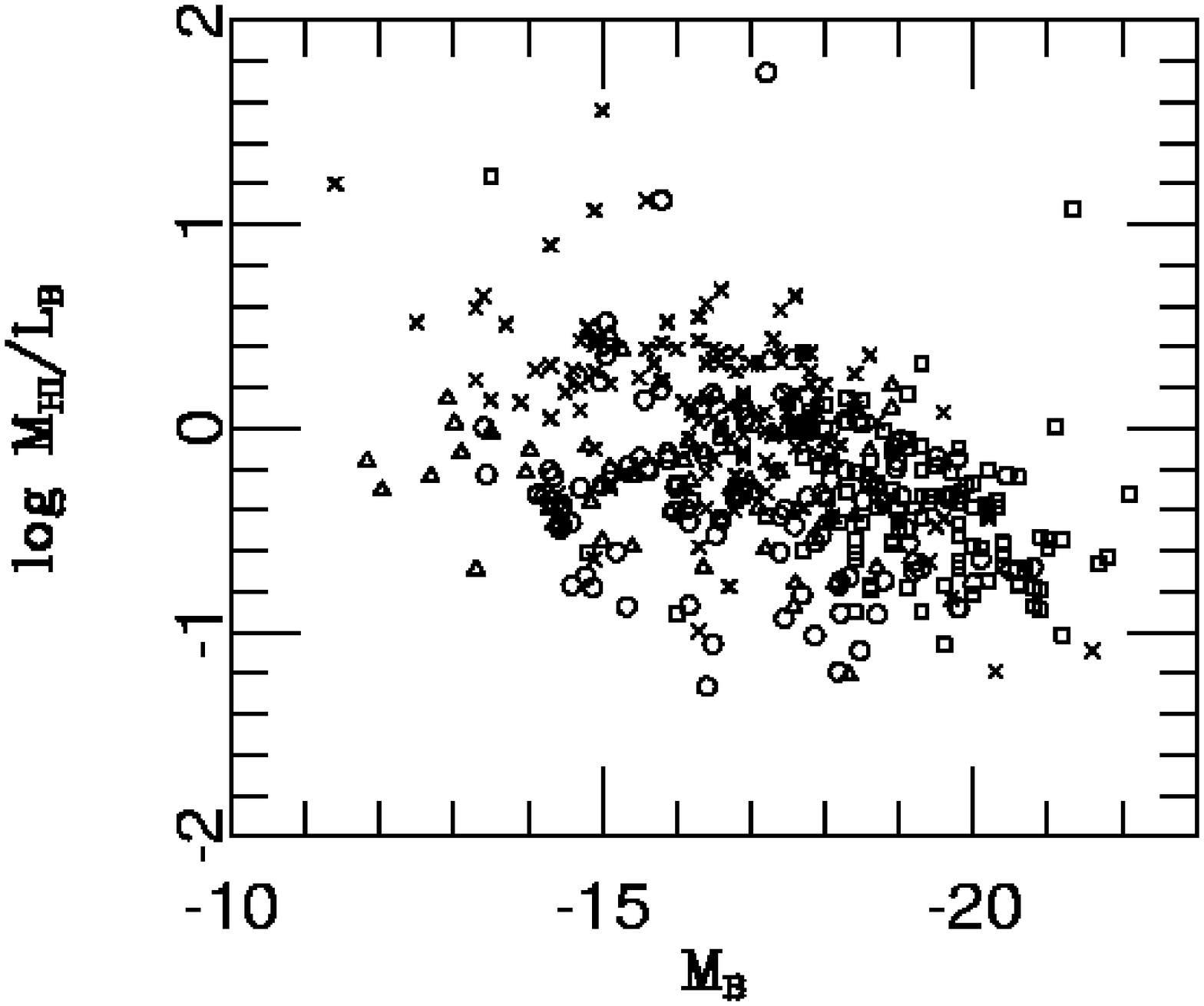}
\narrowcaption[]{The hydrogen mass per B luminosity as function of absolute B 
magnitude for four different types of galaxies: BCGs (circles), dIrr 
(triangles), LSB galaxies (crosses) and late type spiral galaxies (squares). The 
data are obtained from different sources in the literature. It is incomplete but 
estimated to be representative.}
\label{mbhilb}
\end{figure}

Blue compact galaxies (BCGs) is a not well defined type as the galaxies are 
selected either from spectroscopic or photometric critera. The general 
properties are high surface brightness, low chemical abundance and a high gas 
mass fraction. They have a wide range of morphologies (Loose \& Thuan 1986). Are 
they bursting? Fig.~\ref{mbhilb} shows L$_B$/$\cal M_{\rm HI}$ vs. M$_B$ of 
different types of gas rich galaxies. The BCG sample is incomplete but 
constitutes a representative part of the nearby sample of starburst dwarfs (Mrk, 
UM, Tololo etc.). We see that there is a continuous distribution towards high 
L$_B$/$\cal M_{\rm HI}$ but that the properties of most BCGs are similar to dIrr 
and late type spirals of similar luminosity, i.e. they are probably not 
bursting. The high surface brightness of the burst could be due to a high column 
density (and a small scalelength, cf. Papaderos et al.~1996 and Salzer et 
al.~2002), perhaps caused by a low angular momentum. Since their gas mass often 
constitutes a major fraction of the total mass (Salzer et al.~2002), the diagram 
shows that starbursts in these galaxies are either shortlived or rare.  

Some BCGs have a $\sim$~tenfold global increase in SFR, i.e they are true 
starbursts. What are their specific properties? There is no strong indication of 
a correlation between SF activity and tidal interactions (Brosch 2004, Hunter 
and Elmegreen 2004). On the other hand BCGs appear to be involved in mergers 
with intense SF more frequently than other dwarfish galaxies (e.g. Gil de Paz 
2004). It could indicate that mergers are important triggers and morphologically 
shortlived. The gas consumption rates are typically shorter than 100 Myr, i.e. 
similar to the dynamical timescale of a merger.

\subsubsection{Ages and masses}

Dynamical mass estimates of BCGs are difficult since the kinematics sometimes 
are quite chaotic due to the mass motions that cause the burst and because of 
the SN winds. To overcome the problem with the stellar winds it becomes 
necessary to use stellar absorption features. The only useful lines for this 
purpose are the Ca~II triplet lines at about 8500 \AA. Not until quite recently 
has this option become accessible (\"Ostlin et al.~2004). The results are very 
promising and will soon help to solve the question regarding the coupling 
between gas and stars and facilitate the detailed analysis of velocity fields 
based on H$\alpha$ (e.g. Marquart et al. 2004).

\begin{figure}[h]
\includegraphics[width=0.5\textwidth]{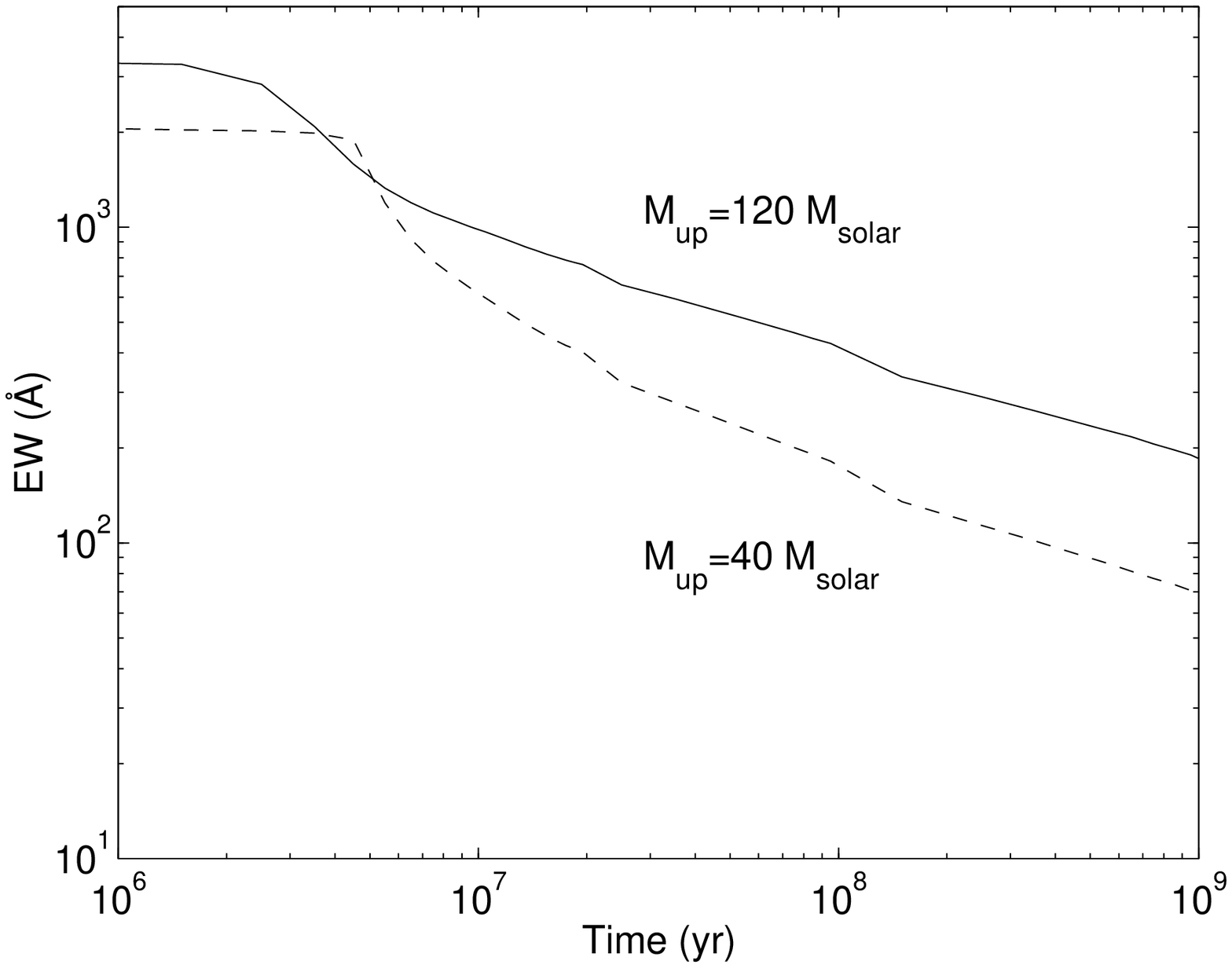}
\narrowcaption[]{The equivalent width of H$\alpha$ in emission from a starburst 
with constant SFR, a Salpeter IMF and 20\% solar metallicity. The predicted 
evolution assuming two different values of the upper mass limit are shown. The 
model is from Zackrisson et al.~(2001).}
\label{ewfig}
\end{figure}

Age and SFR are often estimated from the H$\alpha$ flux, the H$\alpha$ 
equivalent width (EW(H$\alpha$)) and broadband photometry. From this the 
'photometric mass' is obtained assuming that the SFR is constant. The age is 
however difficult to determine, even if we assume that the SFR is constant. In 
such a case, EW(H$\alpha$) is a function of the IMF and age. The IMF slope in 
starbursts seems to be well constrained in the intermediate stellar mass range 
(Elmegreen 2004) but not so well for high masses. Fig \ref{ewfig} shows the 
predicted EW(H$\alpha$) for two values of the upper mass limit, 40 and 120 solar 
masses. It can be seen that the predicted ages differ with a factor of 5-10 over 
a large age range. There is also an observational problem in that intense 
starbursts may have huge Str\"omgrenspheres from which the H$\alpha$ emission 
may be lost due to a limited aperture size. The uncertainty in the determination 
of the widely used b parameter (b = SFR/$<$SFR$>$) obviously must be quite high, 
in particular if we consider the poorly constrained SF history. 

For BCGs there seems to be a simple way to account for the SF history reasonably 
well. It is based on a two component model of the galaxy consisting of a 
starburst superposed on a host galaxy with an exponential luminosity profile. If 
photometric masses are applied to this model we find that there a fairly tight 
correlation between mass and central velocity dispersion (\"Ostlin et al. 2001), 
indicating that this simple model is quite successful.

A very useful method to determine the past starburst activity in a galaxy is 
based on its rich system of super star clusters and globular clusters (GCs). The 
GC IMF is Salpeter-like and their stellar content is coeval. This makes them 
quite reliable as standard clocks and optical/near-IR photometry and 
spectroscopy can be used to determine their ages and trace past star formation 
history thereby identifying bursts (e.g. \"Ostlin et al.~2003; de Grijs et 
al.~2004). 

The best method to derive the ages is from colour-magnitude diagrams of the 
stellar population, but most starburst galaxies are to distant to make this 
method feasible. The few results available give no support for strong shortlived 
bursts separated by quiescent periods (Annibali et al.~2003, Schulte-Ladbeck et 
al.~2001). A similar conclusion was reached by Westera et al.~(2004) in a study 
of 200 HII galaxies based on stellar absorption features. Taking the previous 
discussion into account, these observations indicate that true starbursts are 
rare rather than shortlived or that they are shortlived but change morphological 
type at or after the burst.

\subsubsection{The starburst host}

It is well established that the luminosity profile of most BCGs can be 
characterised by a burst superposed on a host galaxy with (mostly) red colours, 
typical of an old stellar population, and a morphology resembling an early type 
galaxy (e.g. Papaderos et al.~1996, Gil de Paz et al.~2004). An attractive 
scenario is that a dE is merging with a gas rich galaxy that triggers the burst. 
Recently it was found that the optical/near-IR colours of the host of luminous 
BCGs at very faint levels has a red excess, difficult to explain with a normal 
IMF and a low metallicity (Bergvall \& \"Ostlin 2002). This problem is discussed 
in the paper by Zackrisson et al.~(2004). It could indicate that a host galaxy 
of special properties is needed to trigger a true starburst.

{}

\end{document}